\documentclass[twocolumn,showpacs]{revtex4}
\newcommand{\num}{\nonumber}

\begin{document}

\title{ All Static Circularly Symmetric Perfect Fluid Solutions of 2+1
Gravity}
\author{Alberto A. Garc\'{\i}a}
 \email{aagarcia@fis.cinvestav.mx}
\author{Cuauhtemoc Campuzano}
 \email{ccvargas@fis.cinvestav.mx}

\affiliation{Departamento~de~F\'{\i}sica,~%
Centro~de~Investigaci\'on~y~de~Estudios~Avanzados~del~IPN\\
Apdo.\ Postal 14-740, 07000 M\'exico DF, MEXICO}

\date{\today}

\begin{abstract}

Via a straightforward integration of the Einstein equations with
cosmological constant, all static circularly symmetric perfect
fluid 2+1 solutions are derived. The structural functions of the
metric depend on the energy density, which remains in general
arbitrary. Spacetimes for fluids fulfilling linear and polytropic
state equations are explicitly derived; they describe, among
others, stiff matter, monatomic and diatomic ideal gases,
nonrelativistic degenerate fermions, incoherent and pure
radiation. As a by--product, we demonstrate the uniqueness of the
constant energy density perfect fluid within the studied class of
metrics. A full similarity of the perfect fluid solutions with
constant energy density of the 2+1 and 3+1 gravities is
established.
\end{abstract}

\pacs{04.20.Jb, 0440}

\maketitle

\section{Introduction}

In the last two decades a number of researches has been developed
in 2+1 gravity: the search of exact solutions, the quantization of
fields coupled to gravity, topological aspects, black hole
physics, and so on. The literature on this respect is extremely
vast. In the framework of exact solutions, from the beginning till
now, the interest has been focused in the search and the study of
physically relevant solutions and models with sources, for
instance: static ${\it N}$--body spaces~\cite{sta,djh1}, static
and stationary metrics coupled to electromagnetic
fields~\cite{dm,ga,cs2,kk1,ct,cg}, scalar---dilaton
fields~\cite{gak,mz}, cosmology~\cite{bbl}, perfect
fluids~\cite{pc,cf,cz}, among others.

The purpose of this paper is the determination of all static
circularly symmetric spacetimes with cosmological constant coupled
to perfect fluids with and without zero pressure surfaces. In this
context, some progress has been previously achieved. Cornish and
Frankel~\cite{cf} derived all universes obeying a polytropic
equation. Ne\-ver\-theless, since Cornish's solutions were derived
for zero cosmological constant, there is no way to determine a
zero pressure surface. Hence these solutions extend to the whole
spacetime, and consequently they are cosmological solutions. On
the other hand, Cruz and Zanelli~\cite{cz} established some
consequences arising from the hydrostatic equilibrium
Oppenheimer--Volkov equation, and derived for this equation a
single solution for constant energy density; it should be pointed
out that the expression of their $g_{tt}$--metric component
presents a misprint, which is corrected in this work. By the way,
we demonstrate here that the perfect fluid solution with constant
energy density is the only conformally flat---in the sense of the
vanishing of the Cotton tensor~\cite{c1899}---circularly symmetric
solution.

In Sec.~II, by a straightforward integration of the Ein\-stein
equations, we derive the general solution for the static
circularly symmetric 2+1 metric with a cosmological constant
coupled to a perfect fluid solution with variable density $\rho$
and pressure $p$.

Sec.~III is devoted to represent all this class of spacetimes in a
canonical coordinate system. For a given equation of state of the
form $p=p(\rho)$, certain particular families of perfect fluid
solutions are derived; as concrete examples, the subcases of
fluids obeying the linear law $p=\gamma\,\rho$, and those fluids
subjected to a polytropic law $p=C\,\rho^{\gamma}$ in details are
derived.

In Sec.~IV, from the Oppenheimer--Volkov equation certain
properties of the studied solutions are established, for instance,
for positive pressure $p$ and positive density $\rho$, obeying a
state equation $p=p(\rho)$, a microscopically stable fluid
possesses a monotonically decreasing energy density, and
conversely.

In Sec.~V, to facilitate the comparison of the interior
Schwarzschild 3+1 solution with cosmological constant, the perfect
fluid 2+1 solution with $\rho={\rm const.}$ is derived. With this
aim in mind, we search for an adequate representation of the
corresponding structural functions and related quantities of these
3+1 and 2+1 spacetimes. A comparison table is presented. Via a
dimensional reduction of the interior Schwarzschild with $\lambda$
solution, the perfect fluid 2+1 solution with constant $\rho$ is
obtained.

Finally, we end with some concluding remarks.

\subsection {\bf Einstein equations for 2+1 static circularly
symmetric perfect fluid metric}

As far as we know, in most of the publications dealing with the
search of perfect fluid solutions in (2+1) gravity, see for
instance~\cite{pc,cf,cz}, the energy--momentum conservation, i.e.,
the Oppenheimer--Volkov equation, has been used as a clue to
obtain the desired results. On the contrary, we prefer to solve
directly the corresponding Einstein equations; in such case the
energy--momentum conservation equations trivially hold.

The line element of static circularly symmetric 2+1 spacetimes, in
coordinates $\{t,r,\theta\}$, is given by
\begin{eqnarray}
ds^2 = -N(r)^2 dt^2 + \frac{dr^2}{G(r)^2}+r^2 d\theta^2.
\label{met12}
\end{eqnarray}
Note that we are using units such that the velocity of light
$c=1$.

\noindent The Einstein equations with cosmological constant for a
perfect fluid energy--momentum tensor $T_{ab}$:
\begin{eqnarray}
G_{ab}&=& R_{ab} - \frac{1}{2} g_{ab}R = \kappa\,T_{ab}-\lambda
\,g_{ab} ,\nonumber\\
T_{ab}&=&( p+\rho)u_a u_b + p\,g_{ab},
  \,u_a =-{N}\delta^t_a,\nonumber\\
R &=&2\kappa\rho -4\kappa p+6\lambda, \label{ee}
\end{eqnarray}
for the metric~(\ref{met12}) explicitly amount to
\begin{eqnarray}
G_{tt}&=& -\frac { N^2}{ 2r} \frac
{dG^2}{dr} = N^2 (\kappa\rho + \lambda),\num\\
G_{rr}&=& \frac{1}{r N} \frac {d{N}}{dr}  =
\frac{1}{G^2} (\kappa p - \lambda),\num\\
G_{\theta \theta}&=& \frac{r^2}{N}\left[G^2 \frac {d^2N}{dr^2}
+\frac{1}{2} \frac {d N}{dr}\frac {dG^2}{dr} \right]\nonumber\\
&=& r^2 (\kappa p-\lambda). \label{Ein222}
\end{eqnarray}

Notice that the combination of the Einstein equations $ r^2\,G^2
G_{rr} -G_{\theta \theta}=0$, for $N(r)\neq\ {\rm {const.}}$,
gives rise to an important equation:
\begin{eqnarray}
&&G \frac {d^{2}N}{d{r}^{2}}+ \frac {dN}{dr} \left (
 \frac {dG}{dr} - \frac{G}{r} \right )= 0, \nonumber \\
&&\longrightarrow
\frac{d}{dr}\left(\frac{G}{r}\frac{dN}{dr}\right)=0, \label{N15}
\end{eqnarray}
which will be extensively used throughout this paper.

\section{2+1 perfect fluid solution with variable $\rho(r)$}

In this section, we derive the most general static circularly
symmetric solution via a straightforward integration of the
Einstein equations with $\lambda$ for a perfect fluid. It is easy
to establish that the structural functions $G(r)$ and $N(r)$ can
be integrated in quadratures.

Integrating the $G_{tt}$--Eq.~(\ref{Ein222}), one arrives at
\begin{equation}
G(r)^2=- \lambda r^2 -2\kappa\int_0^r{r \rho(r)dr} \equiv\, C -
\lambda r^2 -2\kappa\int^r{r \rho(r)dr}, \label{Gc}
\end{equation}
where $C$ is an integration constant in which we have incorporated
the constant value of the integral at the lower integration limit
$r=0$, thus the remaining integral depends on the upper
integration limit $r$; we use the $r$--notation for the upper
integration limit as well as to denote the integration variable.
This convention will be used hereafter. From the second relation
of Eq.~(\ref{N15}), one obtains
\begin{equation}
\frac{dN}{dr}=n_1\frac{r}{G(r)},
\label{der}
\end{equation}
therefore
\begin{equation}
N(r)=n_1\,\int_0^r{\frac{r}{G(r)}dr}\equiv\,n_0+n_1\int^r{\frac{r}
{G(r)}dr}.
\label{Ncr}
\end{equation}
The evaluation of the pressure $p(r)$ yields
\begin{equation}
\kappa p\,(r)=\frac{1}{N(r)}\left[n_1\,G(r)+\lambda\,N(r)\right].
\label{pr}
\end{equation}

The metric~(\ref{met12}), with $G(r)$ from Eq.~(\ref{Gc}), and
$N(r)$ from Eq.~(\ref{Ncr}), determines the general static
circularly symmetric 2+1 solution of the Einstein equations
(\ref{Ein222}) with $\lambda $, positive or negative, for a
perfect fluid, characterized by a pressure given by
Eq.~(\ref{pr}), and an arbitrary density $\rho(r)$. The
fluid--velocity is aligned along the time--like Killing direction
$\partial_t$. In the derivation of the obtained solutions no
positivity conditions were imposed, thus these fluids allow for
negative $p$ and $\rho$. Nevertheless, to deal with realistic
matter distributions one has to impose positivity conditions on
the density, $\rho>0$, and the pressure, $p>0$, requiring
additionally $\rho>p$.

For a finite distributed fluid, the pressure $p$ becomes zero at
the boundary, say $r=a$; this value of the radial coordinate $r$
is determined as solution of the equation $p(r)=0$.

For non--vanishing cosmological constant, assuming that the values
of the structural functions at the boundary $r=a$ are $N(a)$ and
$G(a)$, the vanishing at $r=a$ of the pressure $p(r)$, given by
Eq.~(\ref{pr}), requires $n_1=-\lambda N(a)/G(a)$, hence
\begin{equation}
\kappa\,p\,(r)=\frac{\lambda}{N(r)\,G(a)}\left[N(r)\,G(a)-N(a)\,G(r)
\right].
\end{equation}

If one is interested in matching the obtained perfect fluid metric
with a vacuum metric with cosmological constant $\lambda$, the
plausible choice at hand is the anti- de Sitter metric, with
$\lambda=-1/l^2$, see Sec.~IV, for which $G(a)= N(a)= \sqrt
{-M_{\infty} + a^2/l^2 }$ at the boundary $r=a$ .

\noindent Incidentally, for a non--zero cosmological constant,
there is no room for dust. The zero character of the pressure
yields the vanishing of the density, and consequently the metric
describes the (anti)--de Sitter spacetime.

For vanishing cosmological constant, the expression of the
pressure (\ref{pr}) is
\begin{equation}
\kappa \,p\,(r)=n_1\frac{G(r)}{N(r)},
\label{pr2}
\end{equation}
from which it becomes apparent that the corresponding solution
represents a cosmological spacetime; there is no surface of
vanishing pressure.

\noindent For vanishing $\lambda$ and zero pressure, the situation
slightly changes: the function $N$ becomes a constant, and the
corresponding  metric can be written as
\begin{equation}
ds^2 = - dt^2 + \frac{dr^2}{C  -2\kappa\int^r{r \rho\,(r)dr}}+r^2
d\theta^2,
\end{equation}
for any density function $\rho$. Of course, the choice of $\rho$
is restricted by physically reasonable matter distributions.

\section{Canonical coordinate system  $\{t,N,\theta \}$}

In this section we show that an alternative formulation of our
general solution can be achieved in coordinates $\{t,N,\theta \}$.
Indeed, from Eq.~(\ref{der}) for the derivative of the function
$N$, in which we are including--without any loss of generality--
the constant $n_1$, (${N}/{n_1}\longrightarrow N, {n_1}\,
t\longrightarrow t$), one obtains
\begin{eqnarray}
\frac{dN}{r}&=&\frac{dr}{G(r)}, \label{der1}
\end{eqnarray}
hence
\begin{eqnarray}
{r}^2&=&{C_0}+\,2\,\int^N{G \,dN}. \label{r2}
\end{eqnarray}
To derive $G$ as a function of the new variable $N$, one uses the
$G_{tt}$---Eq.~(\ref{Ein222}) in the form of
\begin{equation}
G\,dG=-(\kappa \,\rho+\lambda)r\,dr=-(\kappa\,\rho+\lambda)GdN,
\end{equation}
therefore, one gets
\begin{equation}
G(N)=C_1-{\lambda}\,N-{\kappa}\,\int^N{\rho\,(N)\,dN}.
\end{equation}
Substituting this function $G$ into the expression (\ref{r2}) for
$r$, one has
\begin{eqnarray}
H(N):=\,{r}^2&=&C_0+\,2\,C_1\,N-\lambda\,N^2\nonumber\\
&-&2\,\kappa\,\int^N {\int^N{\rho\,(N)\,dN}dN}.
\label{hr4}
\end{eqnarray}
Finally, our metric in the new coordinates $\{t,N,\theta\}$
amounts to
\begin{equation}
ds^2 = -N^2 dt^2 + \frac{dN^2}{ H(N)}+H(N)\, d\theta^2,
\label{met12N}
\end{equation}
which is characterized by pressure
\begin{eqnarray}
p\,(N)=\frac{C_1}{\kappa}\frac{1}{N}-\frac{1}{N}\,\int^N{\rho\,(N)\,dN},
\label{r4}
\end{eqnarray}
and an arbitrary energy density $\rho(N)$ depending on the
variable $N$; for physically conceivable solutions, both functions
$p$ and $\rho$ have to be positive.

The metric (\ref{met12N}) together with the function $H$ from
(\ref{hr4}) give an alternative representation of our general
solution. This representation will be used to derive particular
solutions for a given state equation of the form $p=p\,(\rho)$. In
this approach the expression of the pressure (\ref{r4}) plays a
central role.

For completeness and checking purposes we include the Einstein
equations for perfect fluid and cosmological constant for metric
(\ref{met12N}), considering the function $H$ as an arbitrary one.
There are only two equations governing the problem namely:
\begin{eqnarray}
G_{tt}&:& \frac{d^2 H}{dN^2}=-2\lambda-2\kappa \rho\,(N)\\
G_{NN}&:& \frac{d H}{dN}=-2\lambda N+2\kappa\,N\, p\,(N).
\end{eqnarray}

It is well known that the vanishing of the Cotton tensor in three
dimensions determines locally the class of conformally flat
spaces, in the same fashion as the vanishing of the Weyl tensor
singles out  conformally flat spaces in higher dimensions. The
evaluation of the Cotton tensor
\begin{equation}
C_{\mu\nu\sigma}:=-L_{\mu[\nu;\sigma]},\,L_{\mu\nu;\sigma}:=
R_{\mu\nu;\sigma}-\frac{1}{4}g_{\mu\nu}\,R_{;\sigma}, \label{pr6}
\end{equation}
yields only two independent non--vanishing components
\begin{eqnarray}
C_{ttN}&=&-\frac{1}{4}\,\kappa\,N^2\,\frac{d\rho}{dN},\nonumber\\
C_{\phi\phi\,N}&=&
-\frac{1}{4}\,\kappa\,g_{\phi\phi}\,\frac{d\rho}{dN}.
\end{eqnarray}

Therefore, from the above relations, we conclude that the perfect
fluid solution with $\rho={\rm const.}$ for static circularly
symmetric spacetimes is unique. This result can be stated as a
theorem: The perfect fluid solution with constant $\rho$ is the
only conformally flat static circularly symmetric spacetime for a
perfect fluid source with or without cosmological constant. It is
noteworthy the similarity of this theorem with the corresponding
one formulated for the interior Schwarzschild metric,
see~\cite{col,agdc}. We shall return to this spacetime in Sec.V.

\subsection{2+1 perfect fluid solutions for a linear law
$p=\gamma\,\,\rho$}

Although in the previous section we provided the general solution
to the posed question of finding all solutions for circularly
symmetric static metric in 2+1 gravity coupled to perfect fluid in
the presence of the cosmological constant, from the physical point
of view, even in this lower dimensional spacetime, it is of
interest to analyze certain specific cases, for instance, the
solution corresponding to a fluid obeying the law $p=\gamma \rho$,
or the more complicated case of a polytropic law $p=C\,\rho^\gamma
$.

The starting point in the present study is the linear relation
between pressure and energy density
\begin{eqnarray}
p\,(N)=\gamma\, \rho\,(N).
\label{p00}
\end{eqnarray}
Substituting $p(N)$ from Eq.~(\ref{r4}) into this relation, one
gets
\begin{eqnarray}
\frac{C_1}{\kappa}-\,\int^N{\rho\,(N)\,dN}=\gamma\,N\,\rho\,(N).
\label{po1}
\end{eqnarray}
Differentiating this equation with respect to the variable $N$,
one obtains
\begin{eqnarray}
\frac{d}{dN}{(N\,\rho)}+\frac{1}{\gamma N}\,(N\,\rho)=0,
\label{po12}
\end{eqnarray}
which has as general integral
\begin{eqnarray}
\rho(N)=C_2\frac{\gamma-1}{\gamma^2}{N}^{-\frac{\gamma+1}{\gamma
}},
\label{po3}
\end{eqnarray}
where $C_2$ is an integration constant. Since we arrived at
$\rho(N)$, Eq.~(\ref{po3}), through differentiation, then one has
to replace the obtained result into the relation (\ref{po1}), or
equivalently into Eq.~(\ref{p00}), to see if there arises any
constraint from it:
\begin{equation}
p\,(N)=\frac{C_1}{\kappa}\frac{1}{N}+\gamma\,\,\rho\,(N)=\gamma\,
\,\rho\,(N)\longrightarrow C_1=0, \label{pr3}
\end{equation}
In such manner, we have established that the constant $C_1$
vanishes. Replacing the function $\rho(N)$ from Eq.~(\ref{po3})
into the expression of $H(N)$, Eq.~(\ref{hr4}), and accomplishing
the integration one arrives at
\begin{eqnarray}
H(N)=C_0-\lambda N^2+2\kappa\,C_2N^{(\gamma-1)/{\gamma}}=r^2.
\label{bar}
\end{eqnarray}

Thus, the metric for a perfect fluid fulfilling a linear state
equation in coordinates $\{t,N,\theta\}$ is given by

\begin{eqnarray}
ds^2 =&-&N^2 dt^2 + \frac{dN^2}{C_0-\lambda N^2+2\kappa\,C_2
N^{(\gamma-1)/{\gamma}}}\nonumber\\
&+&(C_0-\lambda N^2+2\kappa\,C_2 N^{(\gamma-1)/{\gamma}})
d\theta^2.
\label{met121}
\end{eqnarray}
To express this solution in terms of the radial variable $r$, one
has to be able to solve the algebraic equation Eq.~(\ref{bar}), in
general a transcendent one, for $N=N(r)$.

The evaluation of the Cotton tensor leads to two independent
non--vanishing components:
\begin{eqnarray}
C_{ttN}&=&C_2 \frac{\kappa}{4}\frac{(\gamma^2-1)}{\gamma^3}\,
N^{-\frac{1}{\gamma}},\nonumber\\
C_{\phi\phi\,N}&=&C_2\frac{\kappa}{4}
\frac{(\gamma^2-1)}{\gamma^3}
\,N^{-\frac{2\gamma+1}{\gamma}}\,g_{\phi\phi}.
\end{eqnarray}
Some examples of physical interest are described by the treated
state equation, for instance: dust, $\gamma=0$, stiff matter,
$\gamma=1$, pure radiation, $\gamma=1/2$, and incoherent
radiation, $\gamma=1/3$. Details the reader may encounter
in~\cite{zn}.

\subsection{2+1 perfect fluid solutions for a polytropic law
$p=C\rho^\gamma$}

This subsection is devoted to the derivation of all solutions
obeying the polytropic law
\begin{equation}
p=C\,\rho^\gamma. \label{pr41}
\end{equation}
Using again the expression of $p\,(N)$ from Eq.~(\ref{r4}), the
above polytropic relation can be written as
\begin{eqnarray}
\frac{C_1}{\kappa}-\,\int^N{\rho(N)\,dN}=C\,N\,{\rho}^{\gamma}(N).
\label{G43}
\end{eqnarray}
Differentiating with respect to $N$, one obtains
\begin{eqnarray}
-\,\rho=C\,\frac{d}{d N}(N\,\rho^{\gamma}),
\label{po15}
\end{eqnarray}
which, by introducing the auxiliary function
$Z=N^{1/\gamma}\,\rho$, can be written as
\begin{eqnarray}
{d}(Z^{\gamma-1})+\frac{1}{C}d(N^{(\gamma-1)/\gamma})=0,
\label{po16}
\end{eqnarray}
therefore
\begin{eqnarray}
{d}\left[\left(\rho^{\gamma-1}+\frac{1}{C}\right)\,N^{(\gamma-1)/
\gamma}\right]=0.
\label{po16}
\end{eqnarray}
Consequently, the general integral of the studied equation becomes
\begin{equation}
\rho= C^{\frac{-1}{\gamma}}\,N^{\frac{-1}{\gamma}}\,\left[B-
C^{\frac{-1}{\gamma}}\,N^{\frac{\gamma-1}{\gamma}}\right]^
{\frac{1}{\gamma-1}},
\label{G441}
\end{equation}
where $B$ is an integration constant. Entering this $\rho$ into
the equation (\ref{pr41}), taking into account that the integral
of the density $\rho$ amounts to
\begin{equation}
\int^N{\rho\,(N)\,dN}=-\int^N{d\,\left[B-
C^{\frac{-1}{\gamma}}\,N^{\frac{\gamma-1}{\gamma}}\right]^
{\frac{\gamma}{\gamma-1}}}, \label{int}
\end{equation}
one arrives at
\begin{equation}
p\,(N)=\frac{n_1}{\kappa}\frac{C_1}{N}+C\,\rho^{\gamma}=C\,
\rho^{\gamma}\longrightarrow C_1=0.
\end{equation}

Considering that the first integral of $\rho$ is given by
(\ref{int}), the expression of the structural function $H(N)$
becomes
\begin{equation}
H=C_0-\lambda\,N^2+
2\,\kappa\int^N\!\!\left[B-C^{\frac{-1}{\gamma}}
\,N^{\frac{\gamma-1}{\gamma}}\right]^{\frac{\gamma}{\gamma-1}}dN=r^2.
\label{G49}
\end{equation}
Notice that the mentioned integral can be expressed in terms of
hypergeometric functions, hence
\begin{widetext}
\begin{equation}
H(N)=C_0-\lambda\,N^2 + 2\,\kappa \, B^{{\gamma}/({
\gamma-1})}\,N\, {\it \large {F}} \left( [{\frac
{\gamma}{\gamma-1}},-{\frac {\gamma}{ \gamma-1}}],[{\frac
{\gamma}{\gamma-1}}+1],{N}^{(\gamma-1)/{\gamma}}{C}^{-1/{\gamma}}
{{B}}^{-1 } \right). \label{G494}
\end{equation}
\end{widetext}

Summarizing, in the case of a polytropic equation of state, the
general solution is given by the metric (\ref{met12N}) with $H(N)$
from (\ref{G49}), and is characterized by energy density and
pressure of the form:
\begin{eqnarray}
p&=& \frac{1}{N}\left[B-
C^{\frac{-1}{\gamma}}\,N^{\frac{\gamma-1}{\gamma}}\right]^
{\frac{\gamma}{\gamma-1}},\nonumber\\
\rho&=& C^{\frac{-1}{\gamma}}\,N^{\frac{-1}{\gamma}}\,\left[B-
C^{\frac{-1}{\gamma}}\,N^{\frac{\gamma-1}{\gamma}}\right]^
{\frac{1}{\gamma-1}}.
\label{G441}
\end{eqnarray}
Incidentally, for zero $\lambda$, the derivation and study of
static circularly symmetric cosmological spacetimes, coupled to
perfect fluids fulfilling the polytropic law was accomplished
in~\cite{cf}, where also have been discussed Robertson---Walker
cosmologies.

The non---vanishing independent components of the Cotton tensor
(\ref{pr6}) are
\begin{eqnarray}
C_{ttN}&=&\frac{1}{4}\kappa
\,N^2\,\frac{d^2}{dN^2}\left[C_1-C^{\frac{-1}{\gamma}}
\,N^{\frac{\gamma-1}{\gamma}}\right]^{\frac{\gamma}{\gamma-1}},
\nonumber\\
C_{\phi\phi\,N}&=&\frac{1}{4}\kappa\,g_{\phi\phi}\,\frac{d^2}{dN^2}
\left[C_1-C^{\frac{-1}{\gamma}}
\,N^{\frac{\gamma-1}{\gamma}}\right]^{\frac{\gamma}{\gamma-1}}.
\end{eqnarray}

These  polytropic fluids contain, amongst others, certain
physically relevant samples: nonrelativistic degenerate fermions,
$\gamma=2$, nonrelativistic matter, $\gamma=3/2$, monatomic and
diatomic gases, $\gamma=7/5$ and $\gamma=5/3$ respectively.

\section{Oppenheimer--Volkov equation}

Although when Einstein equations have been fulfilled, the
energy-momentum conservation law trivially holds, it is of
interest to establish certain properties arising from the
Oppenheimer--Volkov equation, see for instance~\cite{cz} in 2+1
gravity. An alternative derivation of this equation consists in
differentiating with respect to $r$ the Einstein
$G_{rr}$--Eq.~(\ref{Ein222}), this yields
\begin{eqnarray}
\kappa\,\,\frac{dp}{dr}&=& \frac{1}{r N}\frac {d N}{dr}\left(\frac
{dG^2}{dr}-\frac{ G^2}{ r}\right) \nonumber\\
&+& \frac{ G^2}{ r\,N}\left (\frac {d^{2}N}{d{r}^{2}} - \frac{1}{
N}(\frac {d N}{dr})^2 \right). \label{N151}
\end{eqnarray}
Substituting the second derivative ${d^{2}N}/{d{r}^{2}}$ from
Eq.~(\ref{N15}), and the first derivative $ {dN}/{dr}$ from the
$G_{rr}$--equation into Eq.~(\ref{N151}), one arrives at the
Oppenheimer--Volkov equation:
\begin{eqnarray}
{\frac{dp}{dr}}= -\frac{r}{ G^2}(\kappa
\,p-\lambda)(\rho+p).
\label{N152}
\end{eqnarray}
At the circle $r=a$ of vanishing pressure $p(a)=0$, the pressure
gradient amounts to
\begin{eqnarray}
\frac{dp}{dr}{\Large \mid}_{r=a}= \frac{\lambda a}{
G(a)^2}\rho(a).
\label{N153}
\end{eqnarray}
Since inside the circle the pressure is positive, $p(r<a)>0$,
hence at the circle $r=a$ the pressure gradient has to be
non--positive, consequently the cosmological constant ought to be
negative, $\lambda=-1/l^2<0$. We shall continue to use $\lambda$
instead of $-1/l^2$, keeping in mind that $\lambda$ is a negative
constant.

The definition of the mass contained in the circle of radius $a$
is given by
\begin{eqnarray}
M:=2\pi{\int_0}^a{\rho(r)\,r\,dr},
\label{N113}
\end{eqnarray}
and since the metric components $g_{rr}=1/G(r)^2$ has to be
positive in the domain of definition of the solution, then there
exists an upper limit for the mass, namely
\begin{eqnarray}
M\leq\,\frac{\pi}{\kappa} (C-\lambda\,a^2).\label{N114}
\end{eqnarray}

Assuming that a state equation $p=p(\rho)$ holds, the matter
content is said to be microscopically stable if $dp/d\rho\geq 0$.
Since Eq.~(\ref{N152}) can be written as
\begin{eqnarray}
{\frac{dp}{d\rho}}= -\frac{r}{ G^2}(\kappa
\,p-\lambda)(\rho+p)/\frac{d\rho}{dr},
\label{N158}
\end{eqnarray}
one concludes that for a microscopically stable fluid with
positive pressure $p$ and positive density $\rho$, this density
occurs to be monotonically decreasing $d\rho/dr< 0$. Moreover, the
physical requirement that the sound speed is less than the
velocity of light imposes an upper limit on $dp/d\rho\leq 1$.

For our general solution in coordinates $\{t,N,\theta\}$,
me\-tric~(\ref{met12N}), from the expression (\ref{r4}) for the
pressure, one establishes
\begin{eqnarray}
{\frac{dp}{d\rho}}=
-\frac{1}{N}(\rho+p)/{\frac{d\rho}{dN}},\label{N159}
\end{eqnarray}
therefore the density is monotonically decreasing $d\rho/dN< 0$ if
the matter is microscopically stable $dp/d\rho\geq 0$, and
conversely.

Moreover, our fluids, fulfilling the state equation
$p=\gamma\rho$, $\gamma>0$, as well as those ones obeying the
polytropic law $p=C\rho^\gamma$, $C>0,\gamma>0$, are
microscopically stable fluids.

\section{2+1 perfect fluid solution with constant $\rho$ }

As we demonstrated in Sec. II, for constant $ \rho$ the Co\-tton
tensor vanishes, and consequently the corresponding conformally
flat space occurs to be unique.

In this section it is shown that one can achieve a full
correspondence of the metrics and structural functions for
constant energy density perfect fluids in 2+1 and 3+1 gravities.
By an appropriate choice of the constant densities  and
cosmological constants, via a dimensional reduction (freezing of
one of the spatial coordinates of the 3+1 spacetime), one obtains
the 2+1 metric structure from the 3+1 solution. To achieve the
mentioned purpose, the conformally flat static spherically
symmetric perfect fluid 3+1 solution with cosmological constant is
presented in a form which allows a comparison with the static
circularly symmetric perfect fluid with $\lambda$--term and
constant $\rho$ of the 2+1 gravity.

In the canonical coordinate system $\{ t,N,\theta\}$, for $\rho=
\rm{const.}$, the metric, the expression of the function $H$,
which on its turn establishes the relation to the radial
coordinate $r$, and the pressure are given by:
\begin{eqnarray}
ds^2&=&-N^2dt^2+\frac{dN^2}{H}+Hd\theta^2,\\
H&=&C_0+2 C_1 N- (\lambda+\kappa \rho_0)N^2=r^2,\\
p&=&-\rho_0+\frac{C_1}{\kappa}\frac{1}{N}.
\end{eqnarray}
This unfamiliar looking solution can be given in terms of the
radial variable $r$ by expressing $N$ as function of $r$, $
N=N(r)$.

Having in mind the comparison of the 2+1 constant $\rho$ perfect
fluid with its 3+1 relative--the Schwarszchild interior solution--
we shall derive it from the very beginning by integrating the
Einstein equations (\ref{Ein222}) in coordinates $\{t,r,\theta\}$.

For $\rho={\rm const.}$, the integral of (\ref{Gc}) gives
\begin{equation}
G(r)=\sqrt {C -  (\kappa\rho + \lambda)r^2 }. \label{G12}
\end{equation}
Substituting $G(r)$ from (\ref{G12}) into (\ref{Ncr}), one obtains
\begin{equation}
N(r)=n_0-\frac{n_1}{\lambda+\kappa\rho}G(r),
\label{N1r}
\end{equation}
which can be written  as $N(r) = C_1 + C_2 \,G(r)$.

The evaluation of pressure $p(r)$ from Eq.~(\ref{Ein222}) yields
\begin{equation}
\kappa p(r) = \frac{1}{(\kappa\rho+\lambda)N(r)} \left [ n_1
\kappa\rho\,G(r)+n_0\lambda(\kappa\rho +\lambda) \right ],
\label{P12}
\end{equation}
this pressure has to vanish at the boundary $r = a$, which imposes
a relation on the constants:
$n_0=-n_1\,{\kappa\rho}G(a)/[\lambda(\kappa\rho+\lambda)]$, where
$G(a)$ is the value of the function $G(r)$ at the boundary, i.e.,
$G(a)$ is equal to the external value for the $G(r)$ corresponding
to the vacuum  solution plus $\lambda$. A similar comment applies
to $N(a)$. Replacing $n_0$ in Eq.~(\ref{N1r}), the function $N(r)$
becomes
\begin{equation}
\label{Nr2} N(r) = -\frac{n_1}{\lambda(\kappa\rho+\lambda)}\left
[\kappa\rho G(a) +\lambda G(r)\right ].
\end{equation}
Evaluating $N(r)$ at $r=a$, one comes to $ n_1=-\lambda
\,N(a)/G(a)$. Consequently, $N(r)$ amounts to
\begin{equation}
N(r) = \frac{N(a)}{ G(a) ( \kappa\rho +\lambda)}\left [
\kappa\rho\, G(a) + \lambda \,G(r) \right ].
\label{N32}
\end{equation}
Substituting $n_0$, $n_1$, and $N(r)$ into (\ref{P12}), one gets
\begin{equation}
p(r) = \rho\,\lambda\,\frac{G(a) - G(r)}{\kappa\rho G(a) + \lambda
G(r)}. \label{pch}
\end{equation}

Summarizing, the studied perfect fluid for the metric
(\ref{met12}) is determined by structural functions $G(r)$ from
Eq.~(\ref{G12}), and $N(r)$ from Eq.~(\ref{N32}), and
characterized by a density $\rho={\rm const.}$, and pressure $p$
given by Eq.~(\ref{pch}).

The $g_{tt}=-N^2$ metric component, with $N$ from $(\ref{N32})$,
corrects the corresponding one, reported in~\cite{cz}.

\subsection{3+1 conformally flat static spherically symmetric
perfect fluid solution}

In this subsection we review the main structure of the interior
perfect fluid solution in the presence of the cosmological
constant $\lambda$--the interior Schwarzschild metric with
$\lambda$--for the $3+1$ static spherically symmetric metric of
the form
\begin{equation}
ds^2 = -{N(r)}^2dt^2 + \frac{dr^2}{G(r)^2}+r^2 \left (d\theta^2 +
{\rm sin^2}\theta d\phi^2 \right).
\label{met1}
\end{equation}
The Einstein equations for a perfect fluid energy--momentum tensor
in four dimensions have the same form of the ones in three
dimensions (\ref{ee}), except for modifications due to the change
of dimensionality, namely, the expression of $R$ now amounts to
$R= -\kappa T +4\lambda =\kappa\rho -3\kappa p+4\lambda$. Because
the corresponding equations can be found in text--books, we do not
exhibit them here explicitly; we include them for reference in the
Appendix.

Since we are interested in conformally flat solutions, we require
the vanishing of the conformal Weyl tensor, which for static
spherically symmetric spacetime leads to the following equation
\begin{equation}
\frac{d}{dr}\left( \frac{G^2-1}{r^2}\right)=0 \longrightarrow
G(r)=\sqrt {1+c_0 r^2}.
\label{G21}
\end{equation}
On the other hand, from the $G_{tt}$--Eq., one arrives at
\begin{equation}
G(r)=\sqrt{1 -\frac{1}{3}(\kappa\rho+\lambda)r^2}, \label{fc}
\end{equation}
therefore, comparing with Eq.~(\ref{G21}), one has $c_0=
-(\kappa\rho+\lambda)/3,\longrightarrow \rho ={\rm const.}$ Hence,
the solution constructed under this condition would correspond to
a perfect fluid with $\rho ={\rm const.}$, named incompressible
fluid by Adler {\it et al.}~\cite{adl}.

\noindent Moreover, from the
Eq.~$(r^2G^2G_{rr}-G_{\theta\theta})=0$, taking into account the
form of the function $G$ from (\ref{fc}), the general expression
of $N(r)$  is
\begin{equation}
N(r)=C_1+C_2G(r).
\label{Nc}
\end{equation}
The evaluation of the pressure $p$, from $G_{rr}$--Eq., yields
\begin{equation}
\label{p0} \kappa\, p\,(r) = \frac{1}{3 N(r)} \left [C_1\,(
2\lambda-\kappa\rho )-3C_2\,\kappa\,\rho\,G(r) )   \right ].
\end{equation}
where $G(r)$ and $N(r)$ are determined in (\ref{fc}), and
(\ref{Nc})
respectively.\\
This result can be stated in the form of a generalization of the
G\"urses and G\"ursey theorem~\cite{gg} to the case of non--zero
$\lambda$: the only conformally flat spherically symmetric static
solution to the Einstein equations with cosmological constant for
a perfect fluid is given by the metric (\ref{met1}) with
structural functions $G(r)$ and $N(r)$ defined respectively by
(\ref{fc}) and (\ref{Nc}). Moreover, replacing in the
metric~(\ref{met1}) $\rm{sin^2\theta}$ by $\rm{sinh^2\theta}$, and
$\theta^2$, one obtains correspondingly the pseudospherical and
flat branches of solutions.

The constants $C_1$ and $C_2$ are determined through the values of
structural functions at the boundary $r=a$, where the pressure
vanishes, $p(r=a)=0$, they occur to be:
\begin{equation}
C_1 = 3\,C_2 \kappa\rho \,\frac{G(a)}{2 \lambda - \kappa\rho},\,\,
C_2 = \frac{N(a)}{2 G(a)}\frac{2\lambda - \kappa\rho}{\lambda +
\kappa\rho},
\end{equation}
where $G(a)$ is the value of the function $G(r)$ at the boundary
$r=a$, i.e., $G(a)$ is equal to the external value of $G(r)$
corresponding to the vacuum plus $\lambda$ solution. A similar
comment applies to $N(a)$. We shall return to this point at the
end of this subsection.

Substituting the expressions of $C_1$ and $C_2$ into
Eq.~(\ref{Nc}), one has
\begin{equation}
N(r) = \frac{N(a)}{2 G(a) ( \kappa\rho +\lambda)}\left [
3\kappa\rho G(a) + (2\lambda - \kappa\rho)G(r) \right ].
\label{N3}
\end{equation}
Replacing $C_1$, $C_2$ and the above expression of $N(r)$ into
(\ref{p0}), one gets
\begin{equation}
 p\,(r) =  \rho (2 \lambda- \kappa\rho)\frac{G(a) - G(r)}
 {3 \kappa\rho G(a) + (2\lambda - \kappa\rho) G(r)}.
\end{equation}

For the external Schwarzschild with $\lambda$ solution, known also
the Kottler solution~\cite{kot}, the functions $N(r)$ and $G(r)$
are equal one to another, $N(r)=G(r)$, namely
\begin{equation}
N(r) = G(r)=\sqrt{1 - \frac{2 m}{r} -
\frac{\lambda}{3}r^2},\,\,\mbox{for $r\geq a$}.
\end{equation}

Evaluating the mass contained in the sphere of radius $a$ for a
constant density $\rho$, one obtains $2m =\kappa\rho a^3/3$,
therefore
\begin{equation}
N(r) = G(r)=\sqrt{1 - \frac{\kappa\rho}{3}\frac{a^3}{r} -
\frac{\lambda}{3}r^2}, \mbox{for $r\geq a$},
\end{equation}
consequently at $r=a$, one has
\begin{equation}
N(a)=G(a)= \sqrt{1-\frac{\kappa\rho+ \lambda }{3} a^2}.
\end{equation}
In the limit of vanishing cosmological constant, $\lambda = 0$,
one arrives at the interior Schwarzschild solution.

\subsection{Comparison table}

A comparison table of perfect fluid solutions with constant $\rho$
in 2+1 and 3+1 gravities is given as:
\begin{widetext}
\begin{center}
Perfect fluid solutions with constant energy density\\[5pt]
\begin{tabular}{|c|c|}
\hline\hline ${3+1-{\rm solution}}$&${2+1-{\rm solution}}$ \\
\hline ${ds^2=-N^2dt^2+\frac{dr^2}{G^2}+r^2(d\theta^2+{\rm
sin}^2\theta d\phi^2)}$ & ${ds^2=-N^2dt^2+\frac{dr^2}{G^2}
+r^2d\phi^2}$ \\
\hline ${{G^2}=1-\frac{1}{3}(\kappa\rho+\lambda)r^2}$ &
${G^2=C-(\kappa\rho+\lambda)r^2}$ \\
\hline
${N=\frac{1}{2(\kappa\rho+\lambda)}\frac{N(a)}{G(a)}[3\kappa\rho
G(a)+(2\lambda-\kappa\rho)G(r)] }$ &
${N=\frac{1}{(\kappa\rho+\lambda)}\frac{N(a)}{G(a)}[\kappa\rho G(a)
+\lambda G(r)]}$ \\
\hline
${p(r)=\rho(2\lambda-\kappa\rho)\frac{G(a)-G(r)}{3\kappa\rho
G(a)+(2\lambda-\kappa\rho)G(r)}}$ &
${p(r)=\rho\lambda\frac{G(a)-G(r)}{\kappa\rho G(a)+\lambda G(r)}}$ \\
\hline\hline ${{\rm Kottler}:}$ & {{\rm anti--de~Sitter}: $\lambda=-1/l^2$} \\
\hline $G(a)^2=1-\frac{2m}{r}-\frac{1}{3}\lambda r^2$; &
$G(a)^2=-M_{\infty}-\lambda a^2$; \\
$2m=\kappa\rho a^3/3$ & $C=\kappa\rho a^2-M_{\infty}>0$ \\
\hline ${N(a)=G(a)}$&${N(a)=G(a)}$ \\
\hline\hline
\end{tabular}
\begin{tabular}{|c|}
${2\lambda_4-\kappa_4\rho_4 \rightarrow
6\lambda_3},{\kappa_4\rho_4\rightarrow
2\kappa_3\rho_3}$\\
\hline ${G_4(r)\rightarrow G_3(r)},{N_4(r)\rightarrow N_3(r)},
{\kappa_4 p_4(r)\rightarrow2 \kappa_3
p_3(r)}$\\
\hline\hline
\end{tabular}
\end{center}
\end{widetext}

Comparing the structure corresponding to perfect fluid solutions
with constant $\rho$ in 3+1 gravity with the structure of the 2+1
perfect fluid solution one arrives at the following correspondence
: $2\lambda_4-\kappa_4\rho_4 \rightarrow 6\lambda_3$,
$3\kappa_4\rho_4\rightarrow 6\kappa_3\rho_3$, which yields
$G_4(r)\rightarrow G_3(r),N_4(r)\rightarrow N_3(r)$, $\kappa_4
p_4(r)\rightarrow2 \kappa_3 p_3(r)$. Remembering that in 2+1
gravity there is no Newtonian limit, the choice of $\kappa_3$ is
free, thus by selecting $\kappa_3$ appropriately one can achieve
that $p_4(r)\rightarrow p_3(r)$ and $\rho_4\rightarrow \rho_3$.

From this comparison table one easily conclude that the 2+1
perfect fluid with constant $\rho$ can be derived from the
Schwarzschild interior metric by a simple dimensional reduction:
freezing one of the spatial coordinates, say $\theta=\pi/2$, in
the 3+1 solution one obtains the corresponding 2+1 spacetime.

Since we accomplished a scaling transformation of the
$r$-coordinate, accompanied with the inverse scaling of the
angular coordinate $\phi$, one may argue that a conical
singularity could arise; one may overcome this trouble by saying
that the angular coordinate should be fixed once one brings the
2+1 metric to the canonical form with $G_3(r)
=\sqrt{1-(\kappa\rho+\lambda)r^2}$.

\section{Concluding Remarks}

In this contribution we have derived all perfect fluid solutions
for the static circularly symmetric spacetime. The general
solution is presented in the standard coordinate system
$\{t,r,\theta\}$, and alternatively, in a system--the canonical
one-- with coordinates $\{t,N,\theta\}$. From the physical point
of view, particularly interesting are those fluids fulfilling the
linear equation of state, $p=\gamma \rho$, as well as those ones
subjected to the polytropic law $p=C\rho^{\gamma}$; both families
are derived in details from our general metric referred to the
coordinate system $\{t,N,\theta\}$. Therefore, the derived
solutions describe, among others, stiff matter, pure radiation,
incoherent radiation, nonrelativistic degenerate fermions, etc.
The constant energy density perfect fluid solution with
cosmological constant of the 2+1 gravity is singled out among all
static circularly spacetimes as the only conformally flat
space---its Cotton tensor vanishes---sharing the conformally
flatness property with its 3+1 counterpart--the Schwarzschild
interior perfect fluid solution with $\lambda$; a comparison table
for these solutions with constant energy density is included.

\begin{acknowledgments}
This work was partially supported by the CONACyT
Grant 38495E and the Sistema Nacional de Investigadores (SNI).
\end{acknowledgments}

\appendix

\section{3+1  Einstein equations with $\lambda$ for perfect
fluid }

The Einstein equations with cosmological constant for perfect
fluids for the metric (\ref{met1}) explicitly amount  to
\begin{eqnarray}
\label{Ein00} G_{tt}&=& -\frac { N^2}{ r^2}\left ( r \frac
{dG^2}{dr}+G^{2}-1 \right )= N^2 (\kappa\rho + \lambda),\num\\
G_{rr}&=& \frac{1}{G^2 N r^2} \left( 2 r G^2 \frac {dN }{dr} - N+
N G^2 \right)
      =\frac{1}{G^2} (\kappa p - \lambda),\num\\
G_{\theta \theta}&=& \frac{r}{N}\left( G^2 \frac
{dN}{dr}+\frac{1}{2} N \frac {d G^2}{dr} + r G^2 \frac
{d^2N}{dr^2} +\frac{r}{2} \frac {d N}{dr}\frac {dG^2}{dr}
\right)\nonumber \\
&=& r^2 (\kappa p-\Lambda),\num\\
G_{\phi \phi}&=&{\rm sin}^2\theta \,\,\,G_{\theta \theta}.
\end{eqnarray}

\end{document}